\documentclass[12pt]{article}
\usepackage{graphicx,graphics,pifont,amssymb}
\usepackage[figuresright]{rotating}

\begin{document}

\def\tg{{\mbox{\tiny$V$}}}
\def\tl{{\mbox{\tiny$L$}}}
\def\thy{{\mbox{\tiny$H$}}}
\def\te{{\mbox{\tiny$E$}}}

\begin{titlepage}

\title{Levitating Drop in a Tilted Rotating Tank - Gallery of Fluid Motion
\author{Andrew White$^1$, David Swan$^2$ \& Thomas Ward$^1$\\
\small{\emph{$^1$Department of Mechanical and Aerospace Engineering, North Carolina State University}} \\
\small{\emph{Raleigh, NC 27695-7910}}\\
\small{\emph{$^2$Department of Nuclear Engineering, North Carolina State University}} \\
\small{\emph{Raleigh, NC 27695-7909}}}}

\date{} \maketitle

\pagestyle{empty}

\begin{abstract}

A cylindrical acrylic tank with inner diameter D = 4 in is mounted such that its axis of symmetry is at some angle $\alpha$ measured from the vertical plane. The mixing tank is identical to that described in [1] The tank is filled with 200 mL of 1000 cSt silicone oil and a 5 mL drop of de-ionized water is placed in the oil volume. The water drop is allowed to come to rest and then a motor rotates the tank about its axis of symmetry at a fixed frequency $\Omega$ = 0.3 Hz. Therefore the Reynolds number is fixed at about $Re \approx $ 5 yielding laminar flow conditions. A CCD camera (PixeLink) is used to capture video of each experiment.\\

Four experiments are highlighted in this video where the only parameter changed is $\alpha$. In the first clip $\alpha$ = 45$^{\circ}$. The drop of water begins to rotate along the wall of the tank initially, but then begins to center itself about the axis of rotation of the liquid near the bottom of the tank. The axis of rotation of the fluid is distinct from the axis of rotation of the tank and exists below the tank axis of rotation [1].\\

In the second clip $\alpha$ = 50$^{\circ}$ and the same behavior is observed.  The drop appears to settle at a point along the liquid axis of rotation, but a little farther from the bottom of the tank when compared to clip 1.  However instead of settling here the drop then begins to transport towards to free surface along the liquid rotational axis. Closer to the free surface the drop deviates from the liquid rotational axis and settles at the shallow end of the free surface for as long as the tank is rotating. Clips 3 and 4 show experiments were $\alpha$ = 55$^{\circ}$ and 60$^{\circ}$, respectively. The same behavior from clip 2 is observed except that with increasing $\alpha$ the time it takes for the drop to reach the free surface decreases.\\

These levitating drop experiments defy intuition; instead of following solid body rotation where the drop would remain at the tank wall, the drop instead approaches the axis of rotation of the liquid even though its density is greater than that of the oil.\\

\noindent [1] Ward, Thomas, Metchik, Asher, 2007. Viscous fluid mixing in a tilted tank by periodic shear. Chemical Engineering Science 62, 6274-6284.

\end{abstract}

\vspace{1.5cm}

\noindent E-mail: tward@ncsu.edu

\end{titlepage}

\end{document}